\documentclass[journal,comsoc]{IEEEtran}

\def\BibTeX{{\rm B\kern-.05em{\sc i\kern-.025em b}\kern-.08em
		T\kern-.1667em\lower.7ex\hbox{E}\kern-.125emX}}
\usepackage{float}
\normalsize
\usepackage{cite}
\usepackage{graphicx}
\usepackage[cmex10]{amsmath}
\usepackage{relsize}
\usepackage{mathtools}

\usepackage{amsthm,amssymb,hhline}
\allowdisplaybreaks
\usepackage{longtable}
\usepackage{algorithmic}
\usepackage{array}
\usepackage{mdwmath}
\usepackage{mdwtab,bm}
\usepackage{cases}
\usepackage{eqparbox}
\usepackage[tight,footnotesize]{subfigure}
\usepackage{stfloats}

\theoremstyle{remark}

\usepackage{algorithm} 
\usepackage{algorithmic} 
\usepackage{caption}
\usepackage{hyperref}

\usepackage{multirow}

		\usepackage{xcolor}

		{}
		
		\newtheorem{Prop}{Proposition}
		
		\newtheorem{Cor}{Corollory}

		\newtheorem{Remark}{Remark}

\begin{document}
			\title{Self-Connected Spatially Coupled LDPC Codes with Improved Termination}

\author{Yihuan Liao,~\IEEEmembership{Member,~IEEE},
Min Qiu,~\IEEEmembership{Member,~IEEE}, and Jinhong Yuan,~\IEEEmembership{Fellow,~IEEE}

	\thanks{The work was partially supported by the Australian Research Council (ARC) Discovery Project under Grant DP220103596, and in part by the ARC Linkage Project under Grant LP200301482.

The authors are with the School of Electrical Engineering and Telecommunications, University of New South Wales, Sydney, NSW, 2052 Australia (e-mail: yihuan.liao@unsw.edu.au; min.qiu@unsw.edu.au; j.yuan@unsw.edu.au)
}
}

			\maketitle

			\begin{abstract}
				This paper investigates the design of self-connected spatially coupled low-density parity-check (SC-LDPC) codes. First, a termination method is proposed to reduce rate loss. Particularly, a single-side open SC-LDPC ensemble is introduced, which halves the rate loss of a conventional terminated SC-LDPC by reducing the number of check nodes. We further propose a self-connection method that allows reliable information to propagate from several directions to improve the decoding threshold. We demonstrate that the proposed ensembles not only achieve a better trade-off between rate loss and gap to capacity than several existing protograph SC-LDPC codes with short chain lengths but also exhibit threshold saturation behavior. Finite blocklength error performance is provided to exemplify the superiority of the proposed codes over conventional protograph SC-LDPC codes.
			\end{abstract}

			\section{Introduction}
			Spatially coupled low-density parity-check (SC-LDPC) codes, appreciating their excellent decoding threshold and error floor, have received a lot of attention \cite{782171, 5613891,5571910,5695130,6589171,7152893,7553579}. The idea of coupling a sequence of identical LDPC block code (LDPC-BC) graphs into a chained ensemble was first proposed in \cite{782171}. Most importantly, it has been shown both numerically and analytically that the belief propagation (BP) decoding threshold of terminated SC-LDPC ensembles can converge to the optimal maximum a posteriori probability (MAP) decoding performance of the underlying LDPC-BC ensembles \cite{5571910,5695130,6589171}. However, the termination of the coupled code chain is performed by low-degree check nodes (CNs) added to the boundaries The rate loss is not negligible when the coupling length is small.
			
			Due to the rate loss, there exists a noticeable gap between the BP decoding threshold and capacity. To close this gap, \cite{6970871,8590739} have designed randomized irregular SC-LDPC ensembles with non-uniform degree distributions to improve the decoding threshold. In general, the optimization of degree distributions is difficult because the density evolution equations are multi-dimensional. Alternatively, one can design the termination of the code chain to reduce the rate loss. \cite{6404682} considered removing the last few rows of the protograph SC-LDPC base matrix. It has been shown that their proposed SC-LDPC ensembles can achieve the same threshold as the conventional SC-LDPC ensembles with less rate loss. Further, \cite{7593071} considered adding irregular variable nodes (VNs) to the CNs at the chain boundary. The degree distribution of these VNs was optimized to reduce the rate loss without degrading the BP threshold too much. However, it remains unclear whether one can further reduce the rate loss while improving the BP threshold at the same time for small chain lengths.

			Recently, connected-chain SC-LDPC codes were introduced in \cite{6364482,6181771,8718022}, which extend the spatial graph coupling phenomenon from coupling individual LDPC-BC graphs in a chain to more general coupled structures. The main idea is to connect several individual SC-LDPC code chains referred to as single-chain ensembles, to form a multi-chain ensemble. Specifically, additional edges are added to connect the termination CNs in one of the sub-chains to the VNs in other sub-chains. As a result, reliable information propagates from several directions rather than just from the ends of a single chain, leading to an improved BP threshold. In \cite{9594212}, we extended the design of multi-chain ensembles for high-order modulation systems. To improve the decoding threshold of the multi-chain ensembles, \cite{9663207} introduced a new ensemble by connecting a number of small single coupled chains to the loop ensembles from \cite{8718022}. Since the design space of multi-chain ensembles is large, there is still more to explore, such as the connect positions and the connecting structures.
			
			Motivated by the above works, this letter presents a new design of protograph SC-LDPC codes to achieve a reduced rate loss and an improved decoding threshold with short chain lengths. Using a regular SC-LDPC protograph as the building block, we first introduce a termination method to reduce the rate loss. On top of this, we add additional edges between VNs and the remaining termination CNs within the chain, which we refer to as self-connection. We demonstrated that the proposed self-connected SC-LDPC ensembles impose improvements in terms of rate loss and BP threshold over conventional SC-LDPC ensembles. In addition, we extend the proposed construction to design multi-chain ensembles to further improve the BP threshold. We show that the proposed ensembles not only achieve a better trade-off between rate loss and gap to capacity than existing protograph SC-LDPC codes with short chain length but also exhibit threshold saturation behavior. Finally, simulation results demonstrate the superior finite-length error performance of the proposed codes.

			\section{Conventional Protograph SC-LDPC Codes}

\begin{figure}[h]
				\centering
				\includegraphics[width=0.9\linewidth]{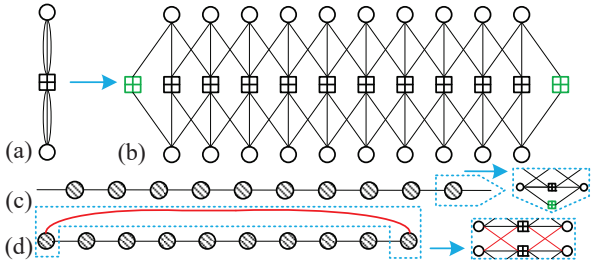}
				\caption{Protpgraph representation of (a) $(3,6)$ regular LDPC-BC, (b) $\mathcal{C}_0(3,6,10,2)$, (c) a simplified representation for $\mathcal{C}_0(3,6,10,2)$, and (d) $\mathcal{T}(3,6,10,2)$.} \label{fig:conventional_scldpc_proto}
			\end{figure}

			In this section, we briefly review the background of protograph SC-LDPC codes \cite{7152893}. A protograph SC-LDPC chain ensemble is constructed by coupling a sequence of $L$ disjoint $(d_{\mathrm{v}},d_{\mathrm{c}})$-regular block protographs into a chain, where the VNs at spatial position $l$ connect the CNs at spatial positions $l,\ldots,l+\omega$ and $l\in[1,L]$. Here, $d_{\mathrm{v}}$, $d_{\mathrm{c}}$, $\omega$, and $L$ represent VN degree, CN degree, coupling width, and coupling length, respectively. The number of VNs and CNs in a $(d_{\mathrm{v}},d_{\mathrm{c}})$-regular LDPC-BC are denoted as $b_{\mathrm{v}}$ and $b_{\mathrm{c}}$, respectively. For illustrative purposes, a $(3,6)$-regular LDPC-BC, as shown in Fig. \ref{fig:conventional_scldpc_proto}(a), is considered as the building block for constructing SC-LDPC codes.
			
			Conventionally, there are two termination methods. The first method adds CNs at the boundaries of a chain, which allows all VNs to share the same degree $d_{\mathrm{v}}$. These additional CNs are called virtual CNs. Furthermore, the boundary areas with regular VNs and irregular CNs are referred to as full termination ends. In addition, the boundary areas without virtual CNs are called non-closure ends. We denote a protograph SC-LDPC chain ensemble by $\mathcal{C}_{t}(d_{\mathrm{v}},d_{\mathrm{c}}, L,\omega)$, where $t\in\{0,1,2\}$ represent the number of open terminations (non-closure ends). With a little abuse of the notation, we sometimes use $\mathcal{C}_t$ to represent $\mathcal{C}_t(d_{\mathrm{v}},d_{\mathrm{c}}, L,\omega)$. For example, a conventional terminated SC-LDPC ensemble $\mathcal{C}_{0}(d_{\mathrm{v}},d_{\mathrm{c}}, L,\omega)$ \cite{7152893} is written as $\mathcal{C}_{0}$. Fig. \ref{fig:conventional_scldpc_proto}(b) shows an example of ensemble $\mathcal{C}_{0}(3, 6, 10, 2)$, where circles and boxes stand for VNs and CNs, respectively.
			
			Note that the CNs at full termination ends are only connected to either 2 or 4 VNs. The code rate $R$ of a terminated SC-LDPC ensemble $\mathcal{C}_{0}$ is given by \cite{7152893}
			\begin{equation}\label{eq:R_0}
				R(\mathcal{C}_{0}) = 1- \frac{(L+\omega)b_{\mathrm{c}}}{Lb_{\mathrm{v}}},
			\end{equation}
			which approaches the code rate of $(d_{\mathrm{v}},d_{\mathrm{c}})$-regular LDPC-BC, $1-b_{\mathrm{c}}/b_{\mathrm{v}}$, as $L \rightarrow \infty$. However, for terminated SC-LDPC with a small $L$, the rate loss is not marginal. Therefore, for a small chain length, it is important to design SC-LDPC ensembles with a reduced rate loss.
			
			Fig. \ref{fig:conventional_scldpc_proto}(c) is a simplified illustration of the protograph representation in Fig. \ref{fig:conventional_scldpc_proto}(b) where each shaded node illustrates a segment consisting of a CN and two VNs. Furthermore, the virtual CNs, represented by green boxes in Figs. \ref{fig:conventional_scldpc_proto}(b-c), are omitted in the simplified representation.
			
			Another well-known termination method is tail-biting, which incurs no rate loss. Use $\mathcal{T}(d_{\mathrm{v}},d_{\mathrm{c}}, L,\omega)$ to represent the tail-biting SC-LDPC ensemble from \cite{7152893}. An example of ensemble $\mathcal{T}(3,6,10,2)$ is shown in Fig. \ref{fig:conventional_scldpc_proto}(d). We notice that all chain positions share the same VN degree and CN degree, which means that the tail-biting structure and its regular LDPC-BC component share the same degree distribution and decoding threshold performance.
			
			\begin{Remark}
				For analytical purposes, the randomized SC-LDPC ensemble is widely considered in the literature, e.g., \cite{5695130,6589171}. However, the randomized code ensemble does not contain any particular protograph-based code ensemble. Moreover, the randomized ensemble does not enjoy a favorable trade-off between rate and threshold as the protograph ensemble \cite{6404671}.
			\end{Remark}
			
			\section{Proposed Protograph SC-LDPC Codes}
			In this section, a new termination approach and a self-connection method are proposed to reduce the rate loss and improve the decoding threshold over conventional SC-LDPC ensembles.

			\subsection{New Termination Methods}
			It is known that the rate loss in the conventional terminated SC-LDPC ensemble $\mathcal{C}_{0}$ comes from adding the virtual nodes in full termination ends. Therefore, reducing the number of virtual nodes at the boundary areas will reduce rate loss. In the following, we present the proposed termination method.

			\begin{figure}[h]
				\centering
				\includegraphics[width=\linewidth]{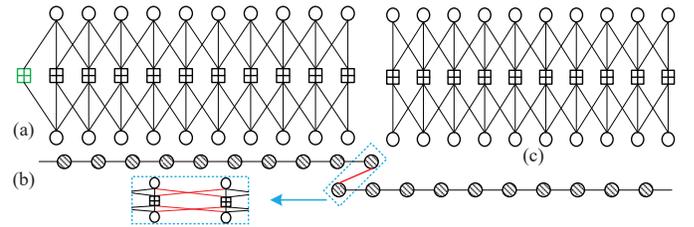}
				\caption{Protograph representation of (a) $\mathcal{C}_1(3,6,10,2)$, (b) a simplified representation of $\mathcal{C}_0(3,6,20,2)$ by connecting two $\mathcal{C}_1(3,6,10,2)$ chains, (c) $\mathcal{C}_2(3,6,10,2)$.} \label{fig:proposed_scldpc_c0_c1}
			\end{figure}
			
			\subsubsection{Single-Side Open SC-LDPC Ensembles}
			Consider coupling $L$ protograph LDPC-BCs via the edge spreading technique of \cite{7152893}. For $\omega>1$ we remove $T\in[1,\omega-1]$ CNs at spatial positions $l\in[1,T]$ or $l\in[L+\omega-T+1,L+\omega]$ from the chain. The resultant chain ensemble contains an end left open with $\omega-T$ CNs with degree less than $d_{\mathrm{c}}$, which we referred to as a \emph{non-closure end}. We call the proposed single-chain SC-LDPC ensemble a single-side open SC-LDPC ensemble $\mathcal{C}_{1}$. The code rate of ensemble $\mathcal{C}_{1}(d_{\mathrm{v}},d_{\mathrm{c}},L,\omega)$ is given by
			\begin{equation}\label{eq:R}
				R(\mathcal{C}_{1}) = 1- \frac{(L+\omega-T)b_{\mathrm{c}}}{Lb_{\mathrm{v}}} \leq 1- \frac{(L+1)b_{\mathrm{c}}}{Lb_{\mathrm{v}}}.
			\end{equation}

			In this work, we set $T=\omega-1$ to minimize the rate loss. As a result, increasing the coupling width no longer increases the rate loss of $\mathcal{C}_{1}$. Compared to $\mathcal{C}_0$, the rate loss reduction under the proposed termination becomes significant when $\omega$ is large. In addition, connecting two single-side open SC-LDPC ensemble $\mathcal{C}_{1}$ forms a conventional terminated ensemble $\mathcal{C}_{0}(d_{\mathrm{v}},d_{\mathrm{c}},2L,\omega)$. An example of ensemble $\mathcal{C}_{1}(3,6,10,2)$ and its relation with $\mathcal{C}_{0}(3,6,20,2)$ are illustrated in Figs. \ref{fig:proposed_scldpc_c0_c1}(a-b).
			
			\subsubsection{Double-Side Open SC-LDPC Ensembles}
			
			To completely eliminate the rate loss of SC-LDPC codes, we remove the CNs at spatial positions $l\in\{1,L+2,\ldots,L+\omega\}$ from $\mathcal{C}_0$. The resultant SC-LDPC ensemble is referred to as a double-side open SC-LDPC ensemble $\mathcal{C}_{2}$. We notice that the ensemble $\mathcal{C}_{2}(d_{\mathrm{v}}, d_{\mathrm{c}}, L, \omega)$ has a structured CN irregularity at boundary areas, which generates reliable information and improves the decoding threshold over the tail-biting ensembles. An example of ensemble $\mathcal{C}_{2}(3,6,10,2)$ is shown in Fig. \ref{fig:proposed_scldpc_c0_c1}(c). However, it is worthwhile to mention that $\mathcal{C}_{2}$ completely eliminates the rate loss at the price of decoding threshold degradation.

			\subsection{Proposed Self-Connection}\label{sec:self}
			On top of the proposed termination, we further propose self-connected SC-LDPC ensembles $\mathcal{S}_{t}$ constructed from adding edges between VNs and low degree CNs within an SC-LDPC ensemble $\mathcal{C}_{t}$. We introduce the following design criteria.
			
			\begin{enumerate}
				\item The maximum allowed CN degree is set to $d_{\mathrm{c}}$.
				\item Edges are added to the CNs at the non-closure ends first, while the edges of the CNs at the full termination ends remain unchanged.
				\item No additional VNs or/and CNs are induced to the chain.
			\end{enumerate}
			
			Constraint 1 guarantees that the CNs used for self-connection are at least as reliable as other CNs in the ensemble after self-connection. Constraint 2 ensures that reliable information can still be generated from the CN at the full termination ends. Moreover, adding edges to the CNs at the non-closure ends can improve the reliability of the information that propagates through them. Even though the degree of some VNs changes, Constraint 3 ensures that the number of CNs and VNs in the protograph of $\mathcal{S}_{t}$ and that of its mother $\mathcal{C}_{t}$ remains the same. Under this condition, the sizes of the parity-check matrices for $\mathcal{S}_{t}$ and $\mathcal{C}_{t}$ after lifting with the same lifting factors are the same. Furthermore, we ensure that the parity-check matrices after lifting are full rank. This requirement is very standard for protograph SC-LDPC codes \cite{7152893,8718022} and can be easily met. Thus, we have $R(\mathcal{C}_{t})=R(\mathcal{S}_{t})$.
			
			We incorporate the above design criteria into determining the connect positions that maximize the decoding threshold via density evolution (DE) \cite[Eqs. (15)-(16)]{7152893}. To this end, we construct the following two ensembles.
			\subsubsection{Self-Connected Single-Side Open SC-LDPC Ensemble}\label{sec:s1}
\begin{figure}[h]
				\centering
				\includegraphics[width=\linewidth]{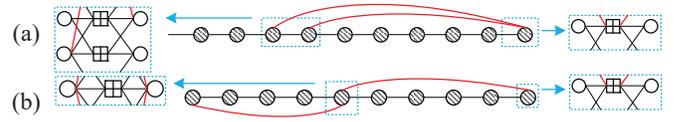}
				\caption{Simplified protograph representation of (a) $\mathcal{S}_1(3,6,10,2)$ and (b) $\mathcal{S}_2(3,6,10,2)$.} \label{fig:proposed_scldpc_s0_s1}
			\end{figure}

			We first connect the CNs at the non-closure end to the VNs located at a distance of $\lfloor L/3 \rfloor$ from the full termination end. Then, the VNs at a distance of $\lfloor L/3 \rfloor+1$ from the full termination end are connected to the CNs at the non-closure end until their degree reaches $d_{\mathrm{c}}$. Intuitively speaking, this allows reliable information propagates from the CNs at the full termination end to the CNs at the non-closure end in two directions. Fig. \ref{fig:proposed_scldpc_s0_s1}(a) shows an example of $\mathcal{S}_{1}(3,6,10,2)$.
			
			\subsubsection{Self-Connected Double-Side Open SC-LDPC Ensemble}
			
			We connect CNs located at the non-closure ends to the VNs at spatial position $\lfloor L/2 \rfloor$, which minimizes the largest spatial distance from the connect position to non-closure ends. In this case, reliable information generated from the connect position propagates to the two boundaries by at most $\lceil L/2 \rceil$ spatial positions. An example of a self-connected double-side open SC-LDPC ensemble $\mathcal{S}_{2}(3,6,10,2)$ is shown in Fig. \ref{fig:proposed_scldpc_s0_s1}(b).

			\subsection{Proposed Multi-Chain Protograph SC-LDPC Ensemble}
			
\begin{figure}[h]
				\centering
				\includegraphics[width=\linewidth]{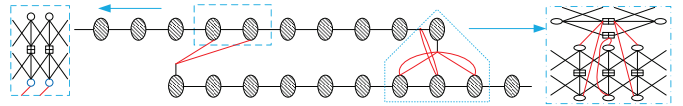}
				\caption{Connecting two $\mathcal{C}_1(3,6,10,2)$ sub-chains to form $\mathcal{M}_1(3,6,10,2)$.} \label{fig:type2}
			\end{figure}

			It has been shown in \cite{8718022} that under short chain lengths, properly designed multi-chain ensembles, such as their proposed loop ensemble $\mathcal{L}(d_{\mathrm{v}}, d_{\mathrm{c}}, L, \omega)$, can achieve a better threshold than the conventional single-chain ensembles while maintaining the linear minimum distance growth. Motivated by these advantages, we use the proposed single-chain ensembles $\mathcal{C}_t(d_{\mathrm{v}}, d_{\mathrm{c}}, L, \omega)$ as the sub-chains to construct multi-chain ensembles $\mathcal{M}_t(d_{\mathrm{v}}, d_{\mathrm{c}}, L, \omega)$. The construction of the proposed multi-chain ensembles follows the design criteria in Sec. \ref{sec:self} whereas the connect positions are determined by maximizing the decoding threshold via DE. The difference is that each sub-chain is not connected to itself but to the other sub-chain. Note that the code rate of $\mathcal{M}_t(d_{\mathrm{v}}, d_{\mathrm{c}}, L, \omega)$ is the same as that for $\mathcal{C}_t(d_{\mathrm{v}}, d_{\mathrm{c}}, L, \omega)$. As an example, we construct $\mathcal{M}_1(3,6,L,2)$ ensemble by connecting two $\mathcal{C}_1(3,6,L,2)$ chains shown in Fig. \ref{fig:type2}.
			
			Observe that the CNs at the full termination end of one sub-chain are connected to the VNs at a distance about $\lfloor L/3 \rfloor$ from the non-closure end at the other sub-chain. Hence, those VNs can benefit from the reliable information propagated from the CNs at the full termination ends of both sub-chains. In addition, a loop is formed such that the CNs at the non-closure ends of both sub-chains are connected via the VNs of the upper sub-chain in Fig. \ref{fig:type2}. In this way, reliable information can propagate to each non-closure end in two directions.

			\begin{figure}[t]
				\centering
				\includegraphics[width=\linewidth]{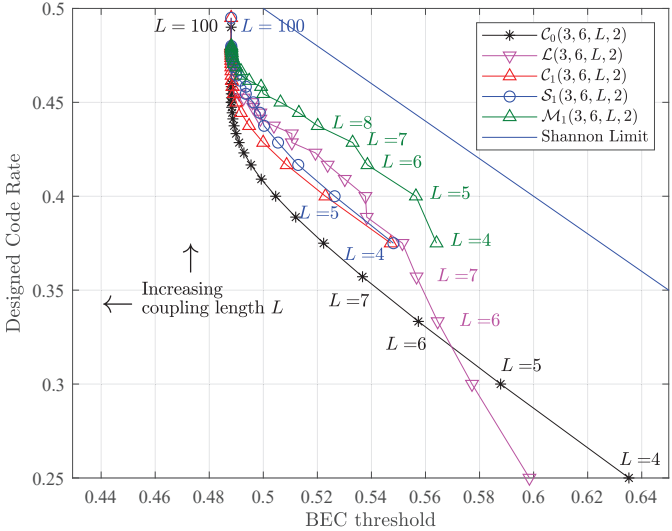}
				\caption{Design rate versus BP threshold on the BEC for various coupled ensembles under different chain lengths.} \label{fig:multi-half-term-threshold}
			\end{figure}

			\section{Decoding threshold Analysis}
			In this section, we use protograph-based DE \cite[Eqs. (15)-(16)]{7152893} to evaluate the BP decoding thresholds of the proposed SC-LDPC ensembles over the binary erasure channel (BEC). Due to space limitations, we omit the detailed DE equations.
			
			\subsection{BP Threshold With a Small Chain Length}
			By using DE, we compute the BP threshold for the proposed ensembles $\mathcal{C}_t(3, 6, L, 2)$, $\mathcal{S}_t(3, 6, L, 2)$, and $\mathcal{M}_t(3, 6, L, 2)$. For comparison purposes, we also include the thresholds of the conventional single-chain regular SC-LDPC ensemble $\mathcal{C}_0(3, 6, L, 2)$ and $\mathcal{T}(3, 6, L, 2)$ from \cite{7152893} and the multi-chain regular SC-LDPC ensemble $\mathcal{L}(3, 6, L, 2)$ from \cite{8718022}. The designed code rates versus the BEC thresholds under different chain lengths are shown in Fig. \ref{fig:multi-half-term-threshold}.

			First, it can be seen that all the proposed ensembles achieve a better trade-off between rate loss and gap to capacity than the conventional single-chain ensemble $\mathcal{C}_0$ with a small chain length. As illustrated in Fig. \ref{fig:multi-half-term-threshold}, for some $L'<L$ the proposed ensembles with chain length $L'$ can achieve both larger rate and threshold than $\mathcal{C}_0$ with chain length $L$. When $L=9$ and compared to the ensemble in \cite[Table I]{6404682} whose threshold is 0.4917, the proposed ensembles $\mathcal{S}_1$ and $\mathcal{M}_1$ achieve larger thresholds, i.e., 0.4987 and 0.5132, respectively, while maintaining the same code rate. When $L\geq25$, the BP threshold of all the proposed ensembles converges to 0.4881. Note also that at $L=40$, the proposed ensembles $\mathcal{C}_1$, $\mathcal{S}_1$, and $\mathcal{M}_1$ achieve the same threshold, i.e., 0.4881, and higher rate, i.e., 0.4875, compared to the ensembles in \cite[Table I]{7593071}. Notably, the proposed ensemble $\mathcal{M}_1$ achieves the best trade-off between rate loss and gap to capacity among all ensembles. By comparing $\mathcal{M}_1$ to the multi-chain ensemble in \cite[Fig. 9]{9663207}, we observe that both ensembles have a similar threshold. It is also interesting to see that the proposed single-chain ensemble $\mathcal{S}_1$ can achieve a slightly better threshold than the multi-chain ensemble $\mathcal{L}$ from \cite{8718022} when $L$ is not small. Since \cite{7152893} has already demonstrated that $\mathcal{C}_0$ outperforms the corresponding uncoupled LDPC-BC in terms of larger threshold and better error performance, we do not repeat the same comparison here.

			\begin{figure}[t]
					\centering
					\includegraphics[width=\linewidth]{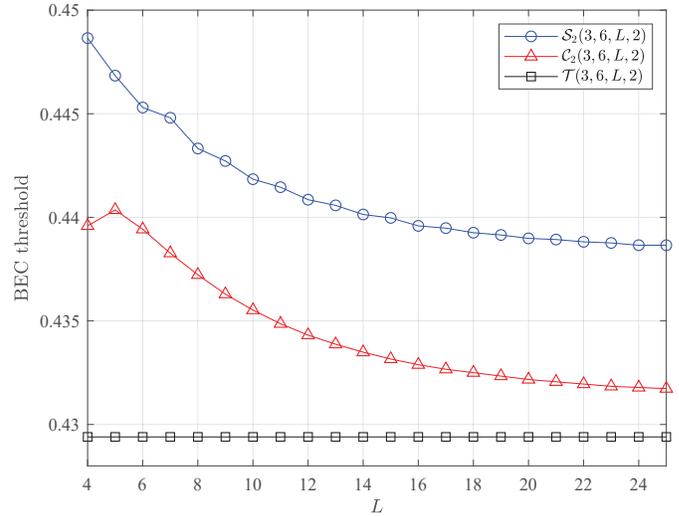}
					\caption{BP threshold for the coupled ensembles without rate loss.}\label{fig:isit_bec}
\end{figure}
			
			Fig. \ref{fig:isit_bec} shows that the BP thresholds of the proposed double-side open SC-LDPC ensembles $\mathcal{C}_2(3,6, L,2)$ and $\mathcal{S}_2(3,6, L,2)$ are larger than tail-biting ensemble $\mathcal{T}(3,6, L,2)$. Interesting, the BP thresholds of $\mathcal{C}_2$ and $\mathcal{S}_2$ decrease with $L$ when $L\geq 5$.
			
			Next, we will show that the proposed ensembles still achieve the same asymptotic performance as $\mathcal{C}_0$ when $L$ goes large.

			\subsection{BP Threshold With a Large Chain Length}\label{sec:BP_thr_L}
			Consider a BEC with erasure probability $\epsilon$. We denote by $P_{j}^{(\ell)}(\mathcal{C})$ the erasure probability at the $j$-th VN of ensemble $\mathcal{C}(d_{\mathrm{v}},d_{\mathrm{c}},L,\omega))$ after $\ell$ iterations, where $j\in \{1,\ldots,V\}$. Its BP decoding threshold is defined as $\epsilon^*(\mathcal{C}(d_{\mathrm{v}},d_{\mathrm{c}},L,\omega))\triangleq \sup\left\{\epsilon>0\left|P_{j}^{(\infty)}(\mathcal{C}) = 0,\forall j\in\{1,\ldots,V\} \right.\right\}$. We have the following proposition on the BP thresholds of $\mathcal{C}_0$ and $\mathcal{C}_1$.
			\begin{Prop}\label{thm1}
				The BP threshold of single-side open ensemble $\mathcal{C}_1$ and the conventional terminated ensemble $\mathcal{C}_0$ satisfies
				\begin{equation} \label{eq:thm}
					\lim\limits_{L\rightarrow\infty}\epsilon^*(\mathcal{C}_1(d_{\mathrm{v}},d_{\mathrm{c}},L,\omega)) = \lim\limits_{L\rightarrow\infty}\epsilon^*(\mathcal{C}_0(d_{\mathrm{v}},d_{\mathrm{c}},L,\omega)).
				\end{equation}
			\end{Prop}
			\begin{IEEEproof}
				First, we need to show that the following holds
				\begin{align} \label{eq:lem_1}
					\epsilon^*(\mathcal{C}_1(d_{\mathrm{v}},d_{\mathrm{c}},L,\omega)) \le \epsilon^*(\mathcal{C}_0(d_{\mathrm{v}},d_{\mathrm{c}},L,\omega)).
				\end{align}
				Note that adding $\omega-1$ virtual CNs to the non-closure end of $\mathcal{C}_1$ results in the conventional terminated ensemble $\mathcal{C}_0$. Since these virtual CNs are used for termination, the reliable information generated by these CNs contributes to lower erasure probabilities for the VNs in $\mathcal{C}_0$ than those in $\mathcal{C}_1$, i.e.,
				\begin{align}\label{eq:pj_comp}
					P_j^{(\infty)}(\mathcal{C}_0) \leq P_j^{(\infty)}(\mathcal{C}_1), \forall j \in\{1,\ldots,V\}.
				\end{align}
				By using \eqref{eq:pj_comp} and the definition of BP threshold at the beginning of Sec. \ref{sec:BP_thr_L}, we get \eqref{eq:lem_1}.

				Next, we show that the following holds.
				\begin{equation} \label{eq:lem_2}
					\epsilon^*\Big(\mathcal{C}_0(d_{\mathrm{v}},d_{\mathrm{c}},2L,\omega)\Big) \le \epsilon^*(\mathcal{C}_1(d_{\mathrm{v}},d_{\mathrm{c}},L,\omega)).
				\end{equation}
				Note that ensemble $\mathcal{C}_0(d_{\mathrm{v}},d_{\mathrm{c}},2L,\omega)$ can be obtained by connecting two sub-chain $\mathcal{C}_1(d_{\mathrm{v}},d_{\mathrm{c}},L,\omega)$ as shown in Fig. \ref{fig:proposed_scldpc_c0_c1}(b), where we highlight the connecting edges in red. Since the decoding schedule does not affect the BP threshold, we can assume that the decoding of $\mathcal{C}_0$ starts from both the termination ends to the middle of the code chain. Compared to the CNs at the non-closure end of $\mathcal{C}_1(d_{\mathrm{v}},d_{\mathrm{c}}, L,\omega)$, the CNs connected by the red edges in $\mathcal{C}_0(d_{\mathrm{v}},d_{\mathrm{c}},2L,\omega)$ have a larger degree and suffer from higher erasure probability. Consequently, the threshold of $\mathcal{C}_0$ is upper bounded by that of $\mathcal{C}_0$ without the red edges. Thus, we have \eqref{eq:lem_2}.
				
				Finally, combining \eqref{eq:lem_1} and \eqref{eq:lem_2} and letting $L \rightarrow \infty$ give \eqref{eq:thm}.
			\end{IEEEproof}
			
			We stress that the proposed ensembles $\mathcal{S}_1$ and $\mathcal{M}_1$ are built from $\mathcal{C}_1$, where the connect positions are determined by optimizing the decoding threshold. Hence, the thresholds of $\mathcal{S}_1$ and $\mathcal{M}_1$ can not be worse than that of $\mathcal{C}_1$. We thus immediately have the following corollary.
			\begin{Cor}\label{cor1}
				The BP thresholds of the proposed ensembles $\mathcal{C}_1$, $\mathcal{S}_1$, and $\mathcal{M}_1$ satisfy
				\begin{align}
					\epsilon^*(\mathcal{C}_1(d_{\mathrm{v}},d_{\mathrm{c}},L,\omega)) &\leq \epsilon^*(\mathcal{S}_1(d_{\mathrm{v}},d_{\mathrm{c}},L,\omega)), \\
					\epsilon^*(\mathcal{C}_1(d_{\mathrm{v}},d_{\mathrm{c}},L,\omega))& \leq\epsilon^*(\mathcal{M}_1(d_{\mathrm{v}},d_{\mathrm{c}},L,\omega)).
				\end{align}
			\end{Cor}
			
			\begin{Remark}		
				We note that the BP decoding threshold of the conventional terminated ensemble $\mathcal{C}_0$ can converge to its MAP decoding threshold \cite{6404671,7152893}. This behavior is known as threshold saturation. With Proposition \ref{thm1} and Corollary \ref{cor1}, we see that the BP thresholds of the proposed ensembles $\mathcal{C}_1$, $\mathcal{S}_1$, and $\mathcal{M}_1$ can also saturate to the MAP threshold of $\mathcal{C}_0$, which approach the BEC capacity asymptotically as $d_{\mathrm{v}}$ and $d_{\mathrm{c}}$ go large.
			\end{Remark}

\begin{figure}[t]
					\centering
					\includegraphics[width=\linewidth]{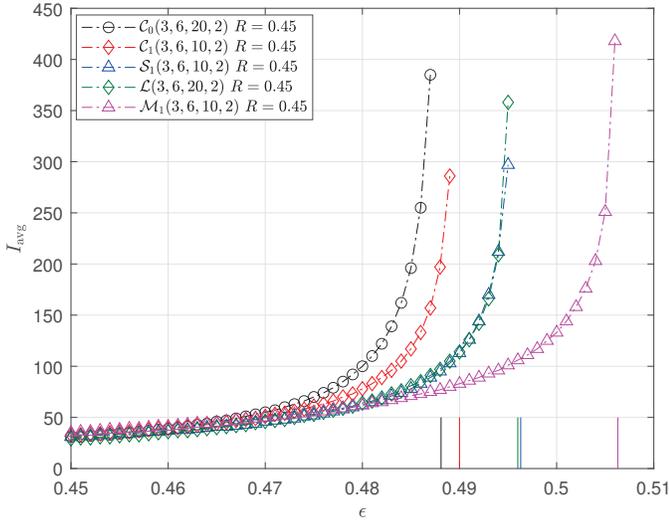}
					\caption{$I_{\rm{avg}}$ versus the BEC
						erasure probability for various ensembles.}\label{fig:complexity_editor}
\end{figure}
				
				\subsection{Decoding Complexity}
				Following \cite{8718022},  we use DE to find the average required iterations per node, denoted by $I_{\rm{avg}}$, for reaching a target BER as the metric for quantifying the decoding complexity. Fig. \ref{fig:complexity_editor} shows $I_{\rm{avg}}$ as a function of the BEC erasure probability $\epsilon$ (dash-dotted line) for reaching a target BER of $10^{-6}$ for five code ensembles with the same design rate 0.45. The parameters of each code ensemble are listed in the legend of Fig. \ref{fig:complexity_editor} and the corresponding decoding threshold is represented by a solid line.
			
				It can be observed that the proposed terminated single-chain ensemble $\mathcal{C}_1$ and $\mathcal{S}_1$ require fewer iterations compared to $\mathcal{C}_0$ \cite{7152893}. It is also interesting to note that the proposed single-chain ensemble $\mathcal{S}_1$ has a similar decoding complexity as that of the multi-chain ensemble $\mathcal{L}$ \cite{8718022}. Finally, the proposed multi-chain ensemble $\mathcal{M}_1$ has the lowest decoding complexity when $\epsilon \geq 0.48$ among all ensembles in Fig. \ref{fig:complexity_editor}. When $\epsilon < 0.48$, the decoding complexity of $\mathcal{M}_1$ is very close to the lowest one.

			\section{Numerical Results}
			
\begin{figure}[t]
					\centering
					\includegraphics[width=\linewidth]{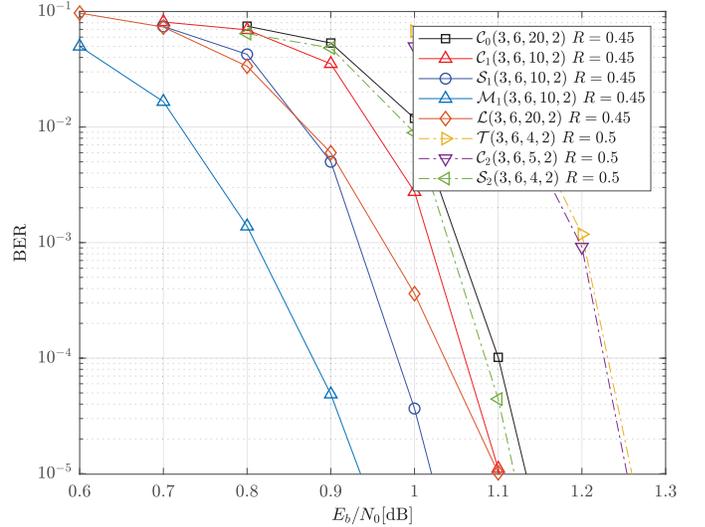}
					\caption{Error performance of various coupled codes on the AWGN channel.}\label{fig:ber_all}
			 \end{figure}

			In this section, we show the bit-error-rate (BER) performance of the proposed codes and compared them with the existing protograph SC-LDPC codes from \cite{7152893,8718022}. For a fair comparison, we consider that all codes have the same total length of $N=80,000$. To achieve the same length, the parity-check matrices of all the investigated protograph single-chain and multi-chain SC-LDPC codes are obtained via lifting their base matrices by $N/(2L)$ and $N/(4L)$, respectively. The error performance is shown in Fig. \ref{fig:ber_all}, where $d_{\mathrm{v}}$, $d_{\mathrm{c}}$, $L$, $\omega$, and code rates are provided in the legend.
			
			At the BER of $10^{-5}$, the proposed protograph SC-LDPC codes $\mathcal{C}_1(3,6,10,2)$ and $\mathcal{S}_1(3,6,10,2)$ effectively improves the decoding performance over $\mathcal{C}_0(3,6,20,2)$ by 0.03 dB and 0.11 dB, separately, while maintaining the same code rate of 0.45. In addition, the proposed SC-LDPC codes without rate loss $\mathcal{C}_2(3,6,5,2)$ and $\mathcal{S}_2(3,6,4,2)$ outperform $\mathcal{T}(3,6,4,2)$ by 0.01 dB and 0.15 dB, respectively. Most notably, the proposed multi-chain SC-LDPC code $\mathcal{M}_1(3,6,10,2)$ achieves the best BER performance by outperforming $\mathcal{C}_0(3,6,20,2)$ and $\mathcal{L}(3,6,20,2)$ by about $0.20$ dB and $0.17$ dB, respectively. It is also interesting to note that the proposed single-chain code $\mathcal{S}_1(3,6,10,2)$ outperforms $\mathcal{L}(3,6,20,2)$. This is in agreement with the DE results in Fig. \ref{fig:multi-half-term-threshold}. Meanwhile, we also observe that the proposed code $\mathcal{S}_2(3,6,4,2)$ outperforms $\mathcal{C}_0(3,6,20,2)$ without sacrificing the code rate.

			\section{Conclusion}
			
			We introduced a new termination method and self-connected structure for protograph SC-LDPC ensembles. For the proposed termination method, we reduced the number of CNs at the full termination ends of a coupled chain to reduce the rate loss due to termination. Further, we proposed a self-connection method to allow reliable information to propagate from several directions to improve the decoding threshold. In addition, we also extended the proposed construction to design new multi-chain SC-LDPC ensembles. We demonstrated that the proposed SC-LDPC ensembles impose noticeable improvements in both rate loss and decoding threshold over several existing SC-LDPC ensembles under short chain lengths. We also showed that the proposed ensembles exhibit threshold saturation behavior. Finite block length error performance is provided to verify the effectiveness of the proposed codes over existing SC-LDPC codes. Future works such as extending the proposed termination and self-connection methods to construct new protograph irregular SC-LDPC codes or multi-chain LDPC codes with more than two coupled chains as in \cite{9663207} are worth further study.

			\bibliographystyle{IEEEtran}
			\bibliography{pubs}

		\end{document}